\documentstyle[aps,multicol]{revtex}

\def\e{\varepsilon}
\renewcommand{\narrowtext}{\begin{multicols}{2} \global\columnwidth20.5pc}
\renewcommand{\widetext}{\end{multicols} \global\columnwidth42.5pc}
\newcommand{\p}{\partial}

\renewcommand{\e}{\varepsilon}
\renewcommand{\H}{{\cal H}}
\renewcommand{\a}{{\bf a}}
\renewcommand{\j}{{\bf j}}

\def\tj{\tilde{j}}

\begin{document}

\title{Universality of transport properties in equilibrium,
Goldstone theorem and chiral anomaly}

\author{Anton Yu. Alekseev$^*$,
 Vadim V. Cheianov$^*$, 
J\"{u}rg Fr\"{o}hlich$^{\dagger}$}

\address{$^*$
Institute for Theoretical Physics, Uppsala University,
Box 803, S-75108, Uppsala, Sweden}

\address{ $^\dagger$
Institut f\"{u}r Theoretische Physik, ETH-H\"{o}nggerberg,
CH-8093, Z\"{u}rich, Switzerland}

\date{March 1998}

\maketitle

{\tightenlines
\begin{abstract}
We study transport in a class of physical systems possessing two 
conserved chiral charges. We describe a relation between universality 
of transport properties of such systems and the chiral anomaly.
We show that the non-vanishing of a current expectation value implies 
the presence of gapless modes, in analogy to the Goldstone theorem. 
Our main tool is a new formula expressing currents
in terms of anomalous commutators. Universality of conductance
arises as a natural consequence of the nonrenormalization of anomalies.
To illustrate our formalism we examine transport properties
of a quantum wire in (1+1) dimensions
and of massless QED in background magnetic field in (3+1) dimensions.

\vskip 0.1cm


\end{abstract}
}

\narrowtext

Transport properties of a physical system are usually linked to complicated 
dynamical processes (such as impurity scattering,
inter-particle interactions {\em etc.}) and, in general, are not universal.
Systems not exhibiting any dissipative processes may, however, 
exhibit some universal transport coefficients that are insensitive 
to changes in the microscopic constitution of the system.
When one encounters a universal transport coefficient one should 
look for a physical principle explaining it's universality.
In the quantum Hall effect, for instance, the universality of the 
Hall conductance can be linked to gauge invariance \cite{L}. 
Another example is superconductivity, where it is the spontaneous 
breaking of the $U(1)$ gauge symmetry 
\cite{Bog} that leads to the vanishing of the longitudinal resistivity
and to the Meissner effect. 
In both examples the conductivity is universal with amazing accuracy, 
which is the result of the existence of a gap in the spectrum of 
bulk charged excitations. Note, however, that an incompressible
quantum Hall system with an edge must support a branch of compressible
edge states, which 
play essential role in understanding Hall quantization.

In this paper we consider a class of physical systems having no gap 
for charged excitations, yet exhibiting
universal transport properties. 
At low energies these systems are assumed to possess two
{\em commuting conserved chiral charges} $Q_L$ and $Q_R$ corresponding
to the particles of left and right chirality. Although our method is not
limited to the electric transport, we consider the electric conductance
as a representative example. The latter is defined as $G=I/V$, where
$I$ is the electric current and 
$eV=\mu_L - \mu_R$ is the difference between the chemical potentials 
$\mu_L$ and $\mu_R$ of the reservoirs of the particles of left and right chirality.
We show that the conductance $G$ is independent of the
dynamics of the system as long as  the charges $Q_L$ and $Q_R$ are conserved.
The principle that protects the universality is the nonrenormalization
of chiral anomalies by interactions \cite{Bar}. To illustrate our
conclusions we examine the transport properties of a quantum wire in
$(1+1)$ dimensions and of massless QED in a background magnetic field
in $(3+1)$ dimensions.

The main tool
we are using is a new formula (\ref{sr}) which 
relates the DC current to the anomalous commutators. In order
to obtain this formula
we use the methods of {\em equilibrium} statistical mechanics ---
an approach that has proven to be effective at the description of 
some transport phenomena in solids \cite{LB,OF,ACF}. 
The only condition imposed on the dynamics of the system is 
the existence of two {\em commuting conserved
charges} $Q_L$ and $Q_R$. 
\begin{equation} \label{assum}
[{\cal H}, Q_L]=[{\cal H}, Q_R]=0 \ , \ [Q_L, Q_R]=0. 
\end{equation}
Here ${\cal H}$ is the Hamiltonian of the theory.
We denote the
conserved Noether currents corresponding to 
the charges $Q_L$ and $Q_R$  by $j^{\mu}_L$ and $j^{\mu}_R$
\begin{equation}
\p_{\mu} j^{\mu}_L=0 \ , \ \p_{\mu} j^{\mu}_R=0.
\end{equation} 
In the examples below 
$j_L$ and $j_R$ are chiral currents corresponding to the fermions of
left and right chirality (in 1+1 or in 3+1 dimensions). The observable 
we are interested in is the electric current $j_e^{\mu}=e(j_L^{\mu} +j_R^{\mu})$ ($e$ is the elementary electric charge).
Henceforth, we will refer to the difference $j_a^{\mu}=e(j_L^{\mu} -j_R^{\mu})$
as to the axial current.

Physically, the conservation of the 
charges $Q_L$ and $Q_R$ means that there is no scattering of the 
particles of left chirality
to the right ones and vice versa.

The conserved charges $Q_L$ and $Q_R$ are conjugate to the chemical 
potentials $\mu_L$ and $\mu_R$ of the external reservoirs of the 
particles of left 
and right chirality.
The thermal state of the system connected to the 
external reservoirs is given
by the density matrix 

\begin{equation} \label{Sig}
\Sigma_{\mu}=e^{-\beta \H_{\mu}} \ , \ 
\H_{\mu}= {\cal H} + \mu_LQ_L +
\mu_R Q_R
\end{equation}
and its transport properties are
described by {\em equilibrium} statistical mechanics.

The continuity equation for the electric current $j_e^{\mu}=(\rho, \j)$ reads

\begin{equation}
\p_\mu j_e^\mu =0.
\end{equation}
In $(d+1)$ dimensions it can be solved in terms of an antisymmetric
tensor field $b$ of rank $d-1$:

\begin{equation}
j_e^\mu = \epsilon^{\mu \nu_1 \dots \nu_d}
\p_{\nu_1} b_{\nu_2  \dots \nu_d},
\end{equation}
where $\epsilon^{\mu \nu_1 \dots \nu_d}$ is Levi-Civita's antisymmetric
tensor.
Note, that the field $b$
is not a physical observable  of the  system. A shift $b \mapsto
b + h$, where the tensor $h$ satisfies 
$\epsilon^{\mu \nu_1 \dots \nu_d}\p_{\nu_1} h_{\nu_2  
\dots \nu_d}=0$,
does not change the physical quantities $j^\mu$
and can be interpreted as a gauge transformation.
This gauge freedom results in the field $b$ only having
$d$ physical degrees of freedom. One may choose a specific
gauge for the potential $b$ which allows to explicitly 
express the physical
degrees of freedom of the field $b$ in terms of a d-vector
field $\a$:  
\begin{equation}
\rho= e \ \nabla \cdot \a \ , \ \j= - e \ \p_t \a.
\end{equation}

In the equilibrium state characterized by the chemical potentials
$\mu_L$ and $\mu_R$ the expectation value of the current is
given by

\begin{eqnarray} \label{j2}
\langle \j(x) \rangle_{\mu}= - e \langle \p_t \a(x) \rangle_{\mu} = 
\frac{ie}{\hbar} \langle [{\cal H}, \a(x)] \rangle_{\mu} = \nonumber \\
=\frac{ie}{\hbar}\ Z_{\mu}^{-1}  {\rm Tr} \ \left(
e^{-\beta H_{\mu}} [{\cal H}_{\mu}, \a(x)] \right) 
-   \nonumber \\
- \frac{ie}{\hbar}\ Z_{\mu}^{-1} {\rm Tr} \ \left(
e^{-\beta H_{\mu}}[\mu_LQ_L +\mu_RQ_R, \a(x)] \right) , 
\end{eqnarray}
where  $Z_{\mu} = {\rm Tr}\ \Sigma_{\mu}$.
At first sight both terms on the r.h.s. of (\ref{j2}) must vanish by
cyclicity of the trace,  because $\H_{\mu}$ and $\mu_LQ_L +\mu_RQ_R$
commute with $\exp(-\beta {\cal H}_{\mu})$. However, we are not allowed 
to use the formula ${\rm Tr}\ [a,b] c = {\rm Tr}\ (abc) - {\rm Tr }\ (bac)$, 
because the triple products are too singular
(not of trace class). 

A more careful analysis shows that only the first trace on the r.h.s
of (\ref{j2}) vanishes.
Indeed, if we regularize the system 
by modifying the Hamiltonian ${\cal H}$ in such a way that a small
spectral gap above the ground state energy is opened then, for a system
in a finite box, 
$e^{-\beta {\cal H}_{\mu}}$ is of trace class,
and the appropriately smeared field $\a(x)$ is bounded by some
function of ${\cal H}_{\mu}$. Then the first trace on the l.h.s.
of (\ref{j2}) vanishes, by cyclicity of the trace. But  this
argument {\em cannot} be applied to the second trace on the r.h.s. of
(\ref{j2}), because,
after regularization, $Q_L$ and $Q_R$ do no longer
commute with ${\cal H}_{\mu}$.

Finally, formula
(\ref{j2}) yields the universal result
\begin{eqnarray} \label{sr}
\langle \j(x) \rangle_{\mu}= - \frac{ie}{\hbar} 
\langle [\mu_LQ_L +\mu_RQ_R, \a(x)] \rangle_{\mu} = \nonumber \\
- \frac{i}{2\hbar} \ (\mu_L - \mu_R) \int dy \ 
\langle [j^0_a(y), \a(x)] \rangle_{\mu} .
\end{eqnarray}
Formula (\ref{sr}) expresses the electric current in terms of the
anomalous commutator
of the time component of the axial current with the field $\a$
solving the continuity equation.

Next, we want to show how nontrivial physical conclusions can be arrived at
by applying formula (\ref{sr}) to concrete physical systems.
Our first example is a one-dimensional interacting electron liquid
(quantum wire). It has long been understood 
\cite{LB} that the conductance
$G=I/V$ (where $I$ is the electric current and $V$
is the voltage drop) of a pure quasi one-dimensional electron 
system must be quantized 
in units of $2e^2/h$ \cite{LB}, {\em i.e.}, 
\begin{equation} \label{un}
G=2n\frac{e^2}{h}, \ n=0,1,2,\dots 
\end{equation}
where $e$ is the elementary electric charge and $h$ is Planck's
constant. The factor of $2$ on the r.h.s. of (\ref{un}) originates
in the spin of electrons, the factor of $n$ corresponds to the number 
of filled energy bands of transversal quantization which form one-dimensional
conducting channels.

It was far from clear, however, how the electron-electron interaction
influenced the conductance. For a long time it was believed that repulsive
electron-electron interactions should suppress the conductance.
Recent experiments \cite{TH} showed that 
the quantization formula (\ref{un}) holds true independently
of the strength of electron-electron interactions in the wire
as long as scattering off impurities is negligible and
the voltage drop $V$ is not very large.

It was argued in \cite{uc} that the nonrenormalization of conductance
by electron-electron interactions was due to the strong influence
of the boundary conditions imposed by the
reservoirs. This idea was supported by calculating the current-current
correlation function of a model one-dimensional system where the
reservoirs were modeled by turning off electron-electron interactions
outside some finite region of the system.
Conductance quantization in quantum wires and Quantum Hall
systems has been compared in \cite{ACF}.

We shall show that a pure quantum wire (with no impurity backscattering)
satisfies the dynamical requirement (\ref{assum})
and the universality of conductance quantization,
as expressed in Eq. (\ref{un}),
follows directly from formula (\ref{sr}).
For simplicity, we shall  consider spinless fermions
and drop the factor of 2 in (\ref{un}). 

The Hamiltonian of a general one-dimensional interacting fermionic system
is given by 
\begin{equation}  \label{int}
{\cal H}= i \hbar v_F \int dx
( \psi^*_L \p_x \psi_L - \psi_R^* \p_x \psi_R) + {\cal H}_{int},
\end{equation}
where $\psi_L$ and $\psi_R$ are left- and right-moving electrons
of the noninteracting model, and ${\cal H}_{int}$ is the interaction
Hamiltonian. It includes higher order terms in $\psi$ corresponding
to electron-electron scattering as well as quadratic terms responsible,
for nonlinearity of dispersion.
We refer to modes created by $\psi^*_L$ and $\psi^*_R$
as to left- and right-movers in spite of the fact that dynamically
they are not necessarily quasi-particles of the interacting model.
One can introduce densities of left- and right-movers
$\psi_L^* \psi_L = n_L \ , \ \psi_R^* \psi_R  =  n_R. $
The total charge density is given by
$\rho=e(n_L + n_R).$
The expression for the electric current density $j$ in terms of $\psi_L$
and $\psi_R$ 
is not universal and depends on the particular form of ${\cal H}_{int}$.

If we assume that the junctions between the one-dimensional system
and the electron reservoirs are adiabatic, the conserved charges
conjugate to the chemical potentials of the reservoirs are equal
to the integrals of $n_L$ and $n_R$:
\begin{equation}
Q_L= \int dx \ n_L \ , \ 
Q_R= \int dx \ n_R.
\end{equation}
We assume that these charges commute with the interacting Hamiltonian
(\ref{int}).

It is convenient to use one-dimensional bosonization formulae
for the Fermi fields $\psi_L$ and $\psi_R$. 

\begin{eqnarray} \label{pb}
\psi^*_L=e^{2\pi i\phi_L} \ , \ \psi_L=e^{-2\pi i\phi_L};
 \nonumber \\
\psi^*_R=e^{-2\pi i\phi_R} \ , \ \psi_R=e^{2\pi i\phi_R}.
\end{eqnarray}
The bosonic fields $\phi_L$ and $\phi_R$ satisfy the
commutation relations
\begin{eqnarray} \label{phicom}
 \ [\phi_{L,R}(x), \phi_{L,R}(y)] = \pm \frac{i}{4\pi}\e (x-y), \nonumber \\
\ [\phi_L(x), \phi_R(y)]= \frac{i}{4\pi}, 
\end{eqnarray}
where  $\e(x-y)=1, x>y;$ and $\e(x-y)=-1, x<y$. The densities of
left- and right-moving particles acquire the form
$n_L= \p_x\phi_L \ , \ n_R=\p_x \phi_R.$
The conserved charges $Q_L$ and $Q_R$
have the following commutation relations with the bosonic fields:

\begin{equation} \label{Qphi}
[Q_L, \phi_L(x)]=\frac{i}{2\pi} \ , \ [Q_R, \phi_R(y)]=-\frac{i}{2\pi} .
\end{equation}
The electric charge density is then given by
\begin{equation}
\rho= e(\p_x \phi_L + \p_x \phi_R) =e \p_x a,
\end{equation}
where $a$ is  the current potential $a=\phi_L+\phi_R$.
We note that all the commutation relations and bosonization
rules listed above only depend on the kinematics of Fermi fields
and are entirely independent of the dynamics of the system.
Our only important dynamical assumption is the commutativity
of the charges $Q_L$ and $Q_R$ with the Hamiltonian of the
interacting system.

Formulae (\ref{sr}) and (\ref{Qphi}) can now be combined to yield
the electrical conductance:

\begin{eqnarray}
\lefteqn{ \hskip -1cm
 \langle j(x) \rangle_{\mu}
=- i\frac{e}{\hbar} \langle [\mu_LQ_L +\mu_RQ_R, a(x)] \rangle_{\mu}=}
\nonumber  \\
&  & \hskip 2cm
 =\frac{e}{h}(\mu_L - \mu_R).
\end{eqnarray}
This concludes the derivation of the universal conductance formula (\ref{un}).

Next, we test formula (\ref{sr}) on a  (3+1)-dimensional example of massless
Dirac fermions coupled to the electromagnetic field. This system is
described by the Lagrangian:
\begin{eqnarray} \label{L3+1}
L= - \frac{1}{4} F^{\mu \nu}F_{\mu \nu} +
\psi_L^* \sigma_L^{\mu} (i\hbar \p_{\mu} - \frac{e}{c} A_{\mu}) \psi_L +
\nonumber \\
+\psi_R^* \sigma_R^{\mu} (i\hbar \p_{\mu} - \frac{e}{c} A_{\mu}) \psi_R ,
\end{eqnarray}
where $\sigma^L_{\mu}=(I,\sigma_k), 
\sigma^R_{\mu}=(I,-\sigma_k)$.
Chiral currents $j_L^{\mu}= \psi_L^* \sigma^L_{\mu}\psi_L$ 
and $j_R^{\mu}=  \psi_R^* \sigma^R_{\mu}\psi_R$ are not conserved
because of the chiral anomaly.
The conservation is recovered upon adding
a Chern-Simons term \cite{4authors}:
\begin{equation}
\tj_{L,R}^{\mu} = j_{L,R}^{\mu} \pm 
\frac{\alpha^2}{8\pi^2 e^2}\
 \epsilon^{\mu \nu \lambda \sigma} A_{\nu} \p_{\lambda} A_{\sigma} .
\end{equation}
where $\alpha$ is the fine structure constant $\alpha=e^2/\hbar c$.
The corresponding charges
\begin{equation}
\label{QLQR}
Q_L= \int d^3x \ \tj_L^0 \ , \ Q_R= \int d^3x \ \tj_R^0
\end{equation}
being conserved, the conjugate chemical potentials $\mu_L$ and $\mu_R$ can be introduced. That $\mu_R$ is different from $\mu_L$ means that the average
density of left-handed particles in the system is different from the average
density of right-handed ones. Such a situation is encountered in
some models of the early Universe \cite{cosmology}.

Our goal is to compute the expectation value of the electric current
$\j$ in the background electromagnetic field $A_{\mu}$
applying formula (\ref{sr}):
\begin{equation} \label{j3+1}
\langle \j(x) \rangle_A = -\frac{ie}{2\hbar} (\mu_L -\mu_R) 
\langle [Q_L - Q_R, \a(x)] \rangle_A .
\end{equation}
The commutators of the densities of the left- and the right-handed fermions  
are given by
\cite{4authors}:
\begin{equation}
[\tj_{L,R}^0(x) ,  \tj_{L,R}^0(y) ] = \pm \frac{i \alpha}{4\pi^2 e}\ 
\p_k \left( B_k(x) \delta(x-y) \right)
\end{equation}
whereas the commutator of the left-handed and the right-handed
current is zero.
Here $B_k = \epsilon^{ijk} \p_i A_j$ is the magnetic field strength.
The commutator of axial and electric charge densities is of the form:

\begin{equation} \label{rr}
[\rho_a(x) , \rho_e(y)] = \frac{ie \alpha}{2\pi^2}\ 
\p_k \left( B_k(x)  
\delta(x-y) \right) .
\end{equation}
Assuming that the commutator of $\rho_a$ with $\a$ is local, one
can remove the divergence on the r.h.s of (\ref{rr}):
\begin{equation} \label{ra}
[\rho_a(x) , a_k(y)] = \frac{i\alpha}{2\pi^2}\
B_k(x) \delta(x-y)+ \dots.
\end{equation}
where the $\dots$ stand for a term of the form of a curl of some vector 
field.
Substituting (\ref{ra}) into (\ref{j3+1}) yields
\begin{equation} \label{jB}
\langle j_k(x) \rangle_A = \frac{\alpha}{4\pi^2 \hbar} \ 
(\mu_L-\mu_R) B_k(x).
\end{equation}
Note that only the first term on the r.h.s. of (\ref{ra}) contributes to the 
current. Another one drops out due to the integration in (\ref{QLQR}). 
The result (\ref{jB}) can be easily verified on the example of a non-interacting 
system in a constant uniform magnetic field, where the single-particle 
picture of \cite{LB} can be implemented. Our derivation implies that formula
(\ref{jB}) holds true when the magnetic field is not necessarily uniform.
In analogue to the previously considered 
quantum wire the formula for the DC current (\ref{jB}) is not affected
by interactions preserving charges $Q_L$ and $Q_R$
\cite{Bar}. Coupling
to a $U(1)$ gauge field in (\ref{L3+1}) is an example of such an 
interaction.

Our next goal is to exhibit a relation between the conductance formula
(\ref{sr}) and the Goldstone theorem. Recall that the Goldstone theorem
states that in a system with spontaneous symmetry breaking
there is a massless
mode (Goldstone boson). At zero temperature the usual proof proceeds as 
follows (see {\em e.g.} \cite{IZ}): assume that
the symmetry group is compact, and denote
the symmetry generators by $L^a$,
\begin{equation} \label{Noether}
L^a = \int dx \ j^a(x),
\end{equation}
where $j^a$ stand for time components of conserved Noether currents.

Spontaneous symmetry breaking
manifests itself in a nonvanishing  expectation value 
\begin{equation} \label{com}
\langle \ [L^a, \Phi(x)] \rangle \neq 0
\end{equation}
of a commutator of a symmetry generator $L^a$ with some operator $\Phi$.
By eq. (\ref{Noether}), (\ref{com}) implies that
\begin{equation} \label{S}
\left( \int_{t=s+\epsilon} dy - \int_{t=s-\epsilon}  dy \right) 
 \ \langle T \left( j^a(t, y) \Phi(s,x) \right) \rangle 
\neq 0 .
\end{equation}
Since the current $j^a_{\nu}$ is conserved, the integration surface
can be deformed into a sphere
of arbitrarily large radius $R$. In order to obtain
a nonvanishing expectation value in (\ref{S}), the correlation
function of $j^a(y)$ and $\Phi(x)$ must decay as $R^{-d}$, where $d$
is the spatial dimension. This implies the existence of massless modes
in the system.

Note that the r.h.s. of the universal conductance formula (\ref{sr})
is exactly of the form (\ref{S}).
The Noether current in this problem is the axial current
$j_a(x)$, the symmetry group is the axial symmetry.
Like in the derivation of the Goldstone theorem, a nonvanishing
expectation value of the current, $\langle j(x) \rangle_{\mu} \neq 0$,
implies the existence of a massless mode in the system.
For instance, in the one dimensional transport problem
this gapless mode is the
density wave described by the field $a(x)$.
The existence of this mode implies
that, in the limit of large distance and low frequency scales, the system
is described by a conformal field theory with a chiral algebra which contains
a $U(1)$ current algebra. In our simple example this conformal field theory
is described by the Luttinger model.

Although we observe a close analogy between the
derivation of the Goldstone theorem and our formula (\ref{sr}),
there is  an important physical difference. 
It is best illustrated
by working out the example of the quantum wire.
There is, in fact, no spontaneous symmetry breaking in the one-dimensional
transport problem. The axial symmetry group of this 
system is $U(1)$. When one introduces an operator $a(x)$ solving
the continuity equation, one decompactifies the axial group from $U(1)$
to $R$. Indeed, under the action of the symmetry generator $Q_L-Q_R$ the 
field $a(x)$ is shifted by a constant,
\begin{equation}
[Q_L-Q_R, a(x)]= \frac{i}{\pi}.
\end{equation}
The  group $U(1)$ has no representations of this type, a constant
cannot be in the same multiplet as a nontrivial field. By introducing
the unphysical field $a(x)$ we effectively  replace  $U(1)$ by
its covering group $R$. Of course, it is not surprising to find that the
expectation value of a constant is nonvanishing. But this fact is not related
to any physical symmetry breaking.
Note, that this situation is special for abelian symmetry groups.

\vskip 0.5cm
{\bf Acknowledgements.}
A.A. and V.C. are grateful to ETH, Z\"{u}rich for hospitality during the period
when this paper was written. We thank J.Adams and J.Mickelsson
for useful discussions.

\widetext

\end{document}